\documentclass[twocolumn,floatfix,10pt,showpacs]{revtex4}
\usepackage[cp1251]{inputenc}
\usepackage[english]{babel}

\usepackage{amssymb}
\usepackage{amsmath}
\usepackage[linktocpage, colorlinks, unicode, citecolor=black, linkcolor=black]{hyperref}
\usepackage{graphicx}

\newcommand{\FFT}{\,\mathrm{FFT}\,}

\begin{document}

\title{Method of numerical simulation of interacting quantum gas kinetics
 on finite momentum lattice}

\author{I. O. Kuznetsov}
\author{P. F. Kartsev}
\affiliation{National Research Nuclear University MEPhI 
(Moscow Engineering Physics Institute), 
115409, Russian Federation, Moscow, Kashirskoe hwy, 31}


\begin{abstract}
We present the efficient and universal numerical method 
for simulation of interacting quantum gas kinetics 
on a finite momentum lattice, based on the
Boltzmann equation for occupation numbers.
Usually, the study of models with two-particle interaction
generates the excessive amount of terms in the equations
essentially limiting the possible system size.
Here we employ the original analytical transformation 
to decrease the scaling index of the amount of calculations.
As a result, lattice sizes as large as 48$\times$48$\times$48
can be simulated, allowing to study realistic problems
with complex interaction models.
The method was applied to the simulation of weakly interacting Fermi
and Bose gases where we calculated the relaxation times
depending on the momentum and temperature.
\end{abstract}

\pacs{71.10.Fd, 72.15.Lh, 67.85.Hj}

\maketitle


\section{Introduction}

There are many problems concerning the time evolution of
complex interacting quantum systems in modern physics.
Examples in solid state physics include 
the behaviour of nonequilibrium charge carriers in semiconductors \cite{Linardy,Chen}
or metals \cite{Rethfeld, kabanov2008},
the relaxation of an excited state in a superconductor \cite{ovchinnikov,Lutchyn},
the condensation of exciton polaritons \cite{Lastras, Berman},
and the dynamics of a Bose-Enstein condensate (BEC)
in a reconfigurable quantum simulator \cite{Sturm}.

To correctly describe 
such physical phenomena as formation of BEC state
\cite{Schulze, Svistunov, Banyai, Straatsma},
relaxation of excited state of electron subsystem 
created by a ultrashort laser pulse \cite{Rethfeld,kabanov2005}
or after absorbing cosmic particle \cite{ovchinnikov},
and many other promising applications,
the detailed theoretical approaches	are required.
Popular analytical approaches include
Boltzmann equation \cite{Wais2018},
nonequilibrium Green functions \cite{Joost2020},
Liouville equation \cite{Tran},
Fermi liquid theory \cite{Aquino},
and others \cite{Kain, Mishonov, kabanov2005}.
Application of purely analytical approaches 
to modern quantum problems, however,
can be difficult due to complexity of the models
and, therefore, numerical calculations are unavoidable \cite{Banyai}.

The Boltzmann equation is efficient for numerical study
of weakly interacting systems not far from equilibrium
\cite{kabanov2008,Wais2018,telenkov2015}.
Here we present the efficient numerical method  
to study the kinetics of quantum systems 
of various statistics 
using the Boltzmann equation
on a finite momentum lattice.
The efficiency is achieved with the special analytical transformation,
which is described in details.
The method allows to take into account 
the energy exchange with thermal bath or nonequilibrium phonons
as well as two-particle interaction,
which usually requires high amount of calculations.
The possibility to take into consideration
 arbitrary single-particle energies 
and level broadening factors 
enables to study extensive set of promising physical problems.
The operation of the method is demonstrated
by the calculation of relaxation times
in weakly interacting Bose and Fermi gases.

\section{Finite momentum lattice}
\label{lattice_chapter}

In this work, we present the numerical method 
to simulate the kinetics of a sufficiently small system,
a finite atomic cluster or nanocrystal.
For certainty, let consider a nanocrystal of
L$\times$L$\times$L atoms with a simple cubic lattice
(generalizations for other lattices are obvious).
The number of points in the reciprocal lattice is also 
L$\times$L$\times$L 
with the step $\Delta k = 2 \pi/La$, 
where $a$ is the lattice constant.

The Hamiltonian of the system is taken in the form:
\begin{eqnarray}
\label{model_hamiltonian}
\hat H = \hat H_1 + \hat H_{\textrm{int}},
\\
\label{H_1}
\hat H_1 = \sum \limits_{\bf k} \varepsilon_{\bf k} \hat n_{\bf k},
\end{eqnarray}
where $\hat H_1$ and $\hat H_{\textrm{int}}$ 
are the single-particle Hamiltonian and interaction part, correspondingly,
$\varepsilon_{\bf k}$ are particle energies,
$\hat{n}_{\bf k}$ is the operator of occupation number.
Index ${\bf k}$ also includes a spin index in the case of fermions.
For complicated problems, additional sorts of particles can be introduced
by relevant terms in the Hamiltonian
\eqref{model_hamiltonian},
for example, phonons to account for the interaction with the lattice.

The kinetics of the system due to interaction is described
by the Boltzmann equation \cite{telenkov}
derived from Fermi's Golden rule:
\begin{equation}
\label{golden_fermi_rule}
\frac{1}{\tau_{\bf k}} = \frac{2 \pi}{\hbar} \sum \limits_{i,f} 
\left<i\right|\hat H_{\textrm{int}}\left|f\right>^2 \delta(E_i - E_f),
\end{equation}
where 
$\left<i\right|\hat H_{\textrm{int}}\left|f\right>$
 are the matrix elements of interaction operator between 
 initial $i$ and final $f$ states
changing the occupation $n_{\bf k}$.
Later on, for simplicity,
we use the time units where $2 \pi / \hbar = 1$.

The application of Boltzmann equation to continual
problems comes down to performing integrations with
the particle density on the energy axis
 \cite{ovchinnikov,kabanov2008}.
In the case of a nanocluster
with finite momentum step, however,
 the numerical summation of the original
expressions should be used.
On the other side,
the number of terms rapidly grows with the lattice size $L$
and sizes $L > 8$ are practically 
untractable \cite{IWOCL2017}.

In this section we present a convenient and efficient
way to extend possible system sizes up to $ L \sim 32 \div 48$
using a special transformation.
This allows simulations of large enough sizes
for subsequent extrapolation
 to the continual limit $L \to \infty$.
As a result, new physical problems can be studied
which could not be explored by the aforementioned analytical methods.

\paragraph{Interaction with the phonon subsystem.}

We begin with a simple case: 
a system of free particles
in the presence of phonons 
(crystal lattice with given temperature $T$):

\begin{equation}
\hat H_{\mathrm{int,phon}} = M_0 \sum \limits_{\bf k q}
{ \hat a^{\dagger}_{\bf k} \hat a_{{\bf k}-{\bf q}} 
\hat b_{\bf q} } + H.c.,
\end{equation}
where $\hat a_{\bf k}$, $\hat a^{\dagger}_{\bf k}$ 
and $\hat b_{\bf q}$, $\hat b^{\dagger}_{\bf q}$
are operators of particles under consideration and phonons, correspondingly,
and $M_0$ is the matrix element of interaction with phonons.
In the case of Fermi particles,
spin index is for simplicity
included in the particle momentum ${\bf k}$.

\begin{figure}
\begin{minipage}{0.4\linewidth}\begin{center}
\includegraphics[width=1.0\linewidth]{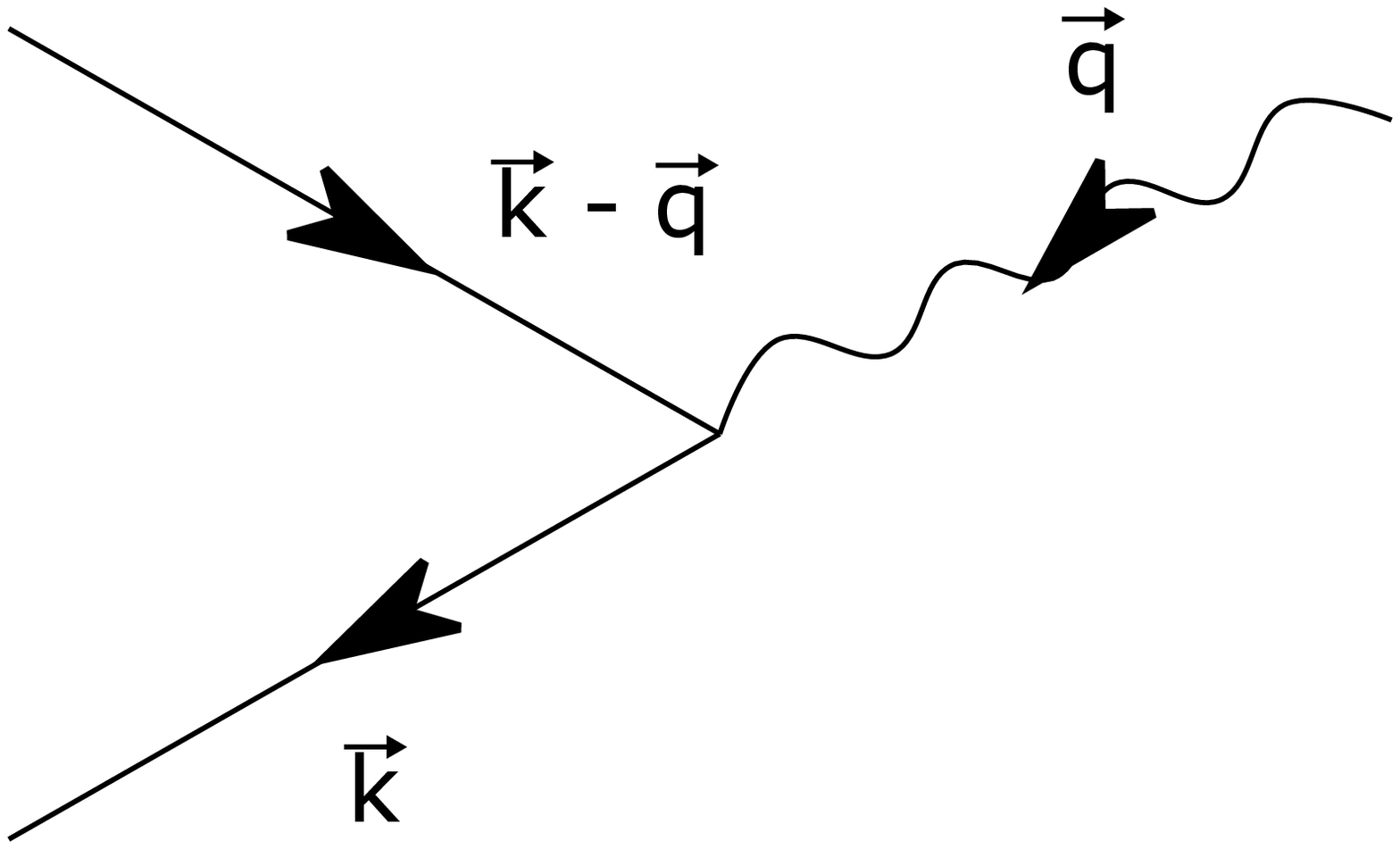}

(a) $P^{(+)}_{\mathrm{in}}$
\end{center}\end{minipage}
\begin{minipage}{0.4\linewidth}\begin{center}
\includegraphics[width=1.0\linewidth]{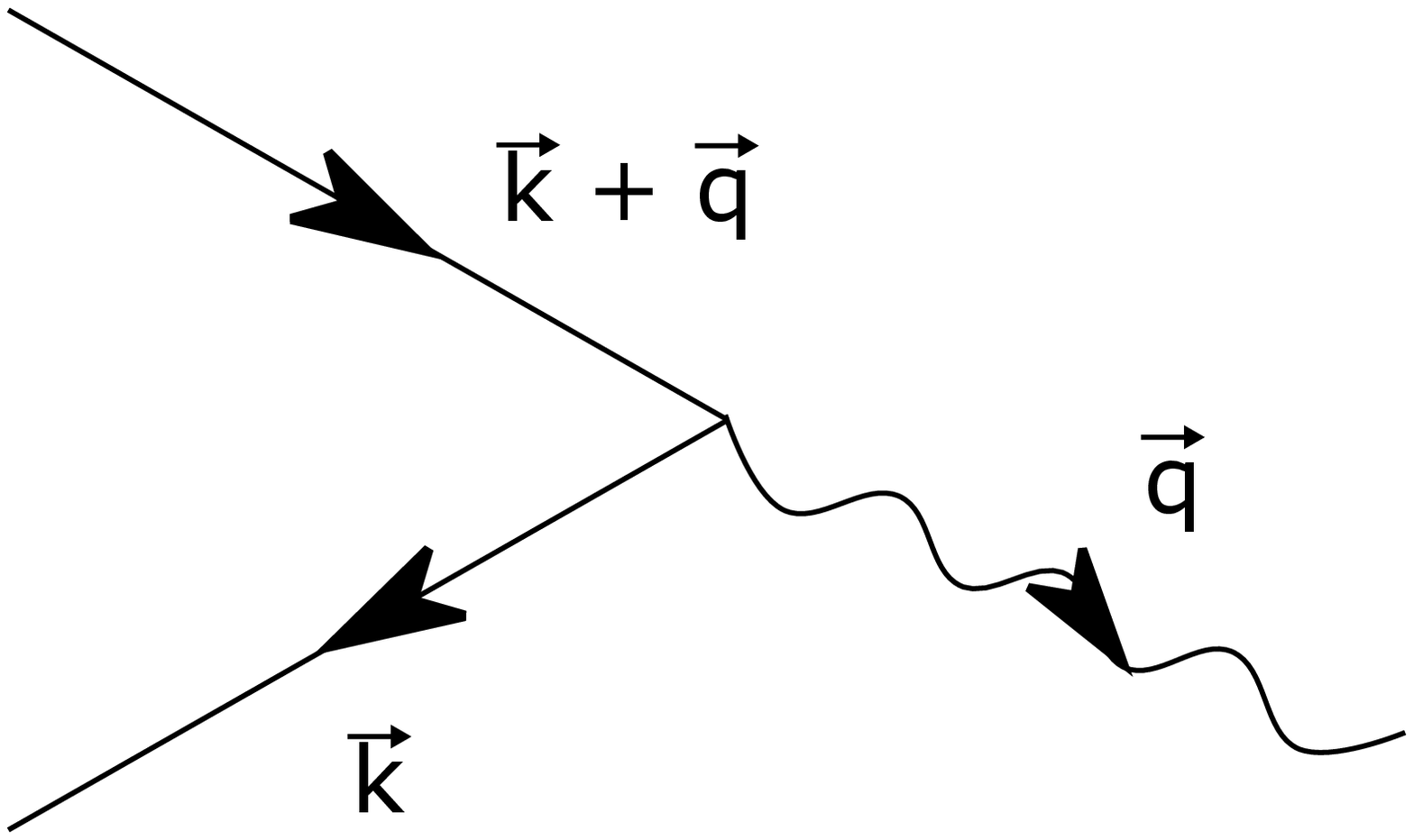}

(b) $P^{(+)}_{\mathrm{out}}$
\end{center}\end{minipage}

\begin{minipage}{0.4\linewidth}\begin{center}
\includegraphics[width=1.0\linewidth]{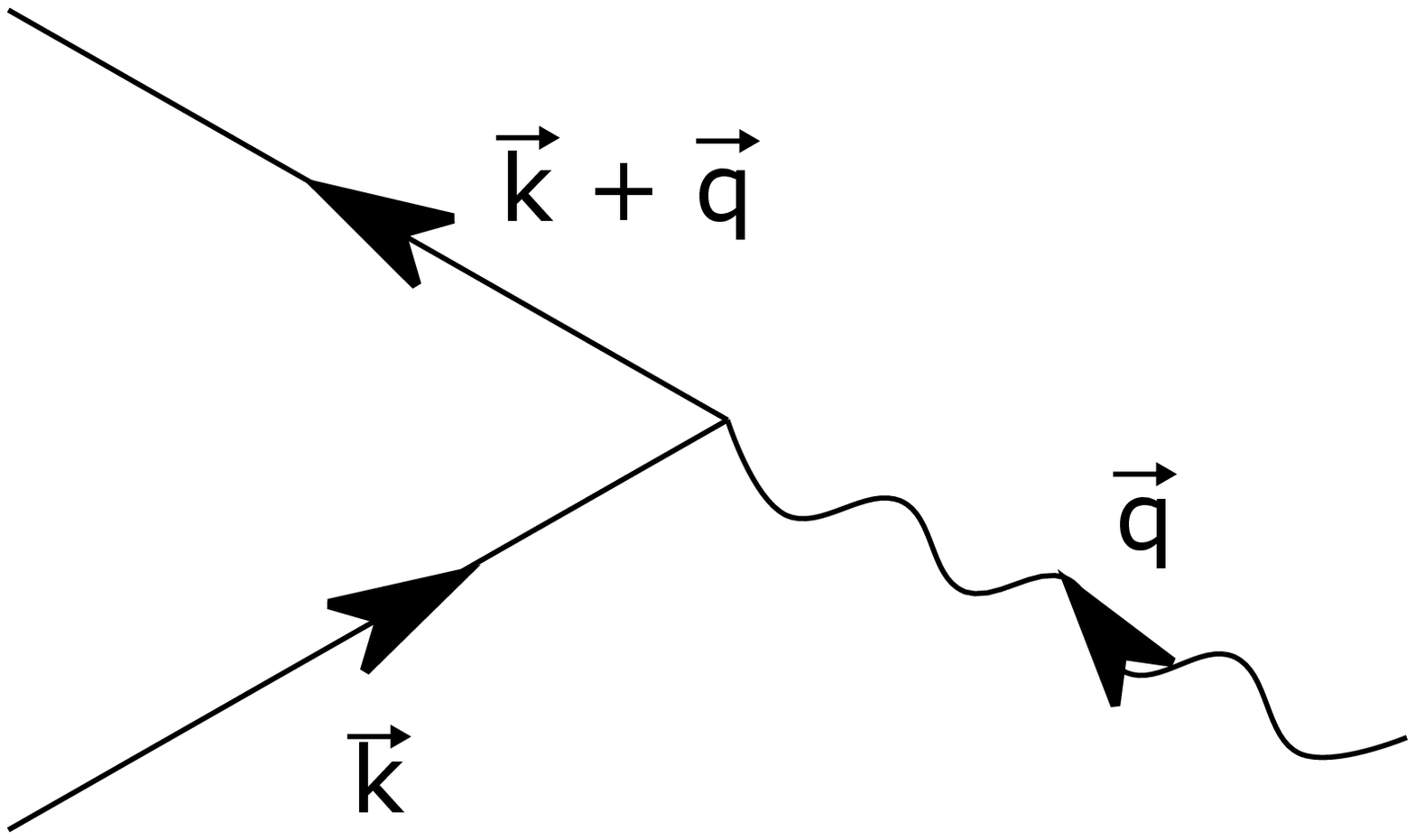}

(c) $P^{(-)}_{\mathrm{in}}$
\end{center}\end{minipage}
\begin{minipage}{0.4\linewidth}\begin{center}
\includegraphics[width=1.0\linewidth]{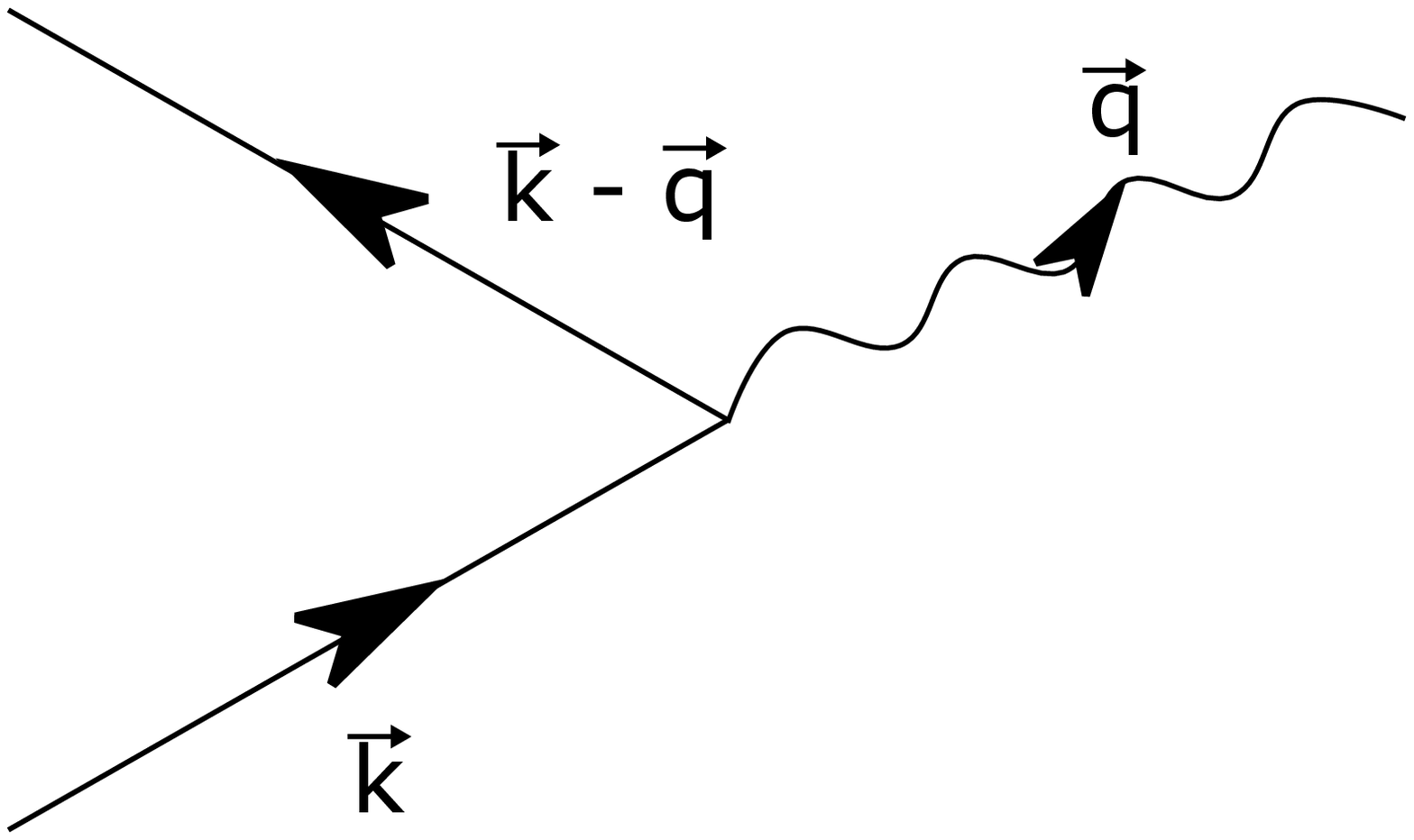}

(d) $P^{(-)}_{\mathrm{out}}$
\end{center}\end{minipage}
\caption{The processes 
for the corresponding terms in Eq. \eqref{dn_dt_phonon}
generated by the interaction with phonons:
adding or removing particle with momentum ${\bf k}$
and phonon with momentum ${\bf q}$.
}
\label{e_ph_figs}
\end{figure}

In Figure \ref{e_ph_figs}, we show 
the processes generated by this type of interaction
that change the number of particles
$n_{\bf k}$ and phonons $n^{\text{(phon)}}_{\bf k}$.
The corresponding terms in the kinetic equations
according to \eqref{golden_fermi_rule} in the case of Bose statistics
have the form:

\begin{eqnarray}
\label{dn_dt_phonon}
\frac{d n_{\bf k}}{d t} = 
P^{(+)}_{\mathrm{in}} +
P^{(+)}_{\mathrm{out}} 
- P^{(-)}_{\mathrm{in}} 
- P^{(-)}_{\mathrm{out}} ,
\\
\nonumber
P^{(+)}_{\mathrm{in}} = M^2_0 \sum\limits_{\bf q}
 (n_{\bf k} + 1) n_{{\bf k}-{\bf q}}  n_{\bf q}^{\text{(phon)}}
F( \delta E ),
\\
\nonumber
P^{(+)}_{\mathrm{out}} = M^2_0 \sum\limits_{\bf q}
 (n_{\bf k} + 1) n_{{\bf k}+{\bf q}} (n_{\bf q}^{\text{(phon)}} + 1)
F( \delta E ),
 \\
\nonumber
P^{(-)}_{\mathrm{in}} = M^2_0 \sum\limits_{\bf q}
 n_{\bf k} (n_{{\bf k}+{\bf q}} + 1 ) n_{\bf q}^{\text{(phon)}}
 F( \delta E ),
 \\
\nonumber
P^{(-)}_{\mathrm{out}} = M^2_0 \sum\limits_{\bf q}
 n_{\bf k} (n_{{\bf k}-{\bf q}} + 1 ) (n_{\bf q}^{\text{(phon)}} + 1)
F (\delta E ),
\end{eqnarray}
where phonon energies are denoted $\tilde \varepsilon_{\bf q}$ 
and $\delta E$ is the correspoding energy difference
for incoming and outcoming particles.
The factor $F (\delta E)$ is introduced 
to account for the finite width of the energy levels.
The four given terms correspond to the 
various types of interaction
with phonon absorption or radiation, as shown in Figure \ref{e_ph_figs}.
In the case of Fermi statistics, $(n+1)$ is replaced with $(1-n)$.

Depending on the problem under study,
the phonon subsystem can be taken into account in two ways:
considering phonons as equilibrium (thermal bath) or nonequilibrium.
For equilibrium phonons, their occupation numbers
are determined by the Bose-Einstein distribution function
$n_{\bf q}^{\text{(phon,0)}} = f( \tilde \varepsilon_{\bf q}, T)$,
here $T$ is the temperature.
In the case of nonequilibrium phonons, the system \eqref{dn_dt_phonon}
is supplemented with similar equations for the phonon numbers,
and finite phonon lifetime
$\tau^\text {(phone)} $ 
related to their decay in the area of consideration
is taken into account:
\begin{equation}
\label{tau_phonon}
P^{\mathrm{(phon)}}_{\mathrm{dec}} = - \frac{1 }{\tau^\text{(phon)}}
  \left( n_{\bf q}^\text{(phon)} -  n_{\bf q}^\text{(phon,0)} \right).
\end{equation}

The interaction with the phonon subsystem is crucial when
the pair interaction is negligible, 
for example, at low particle density.

\paragraph{Two-particle interaction.}
However, the interaction with phonons alone is often not enough
to correctly describe kinetic phenomena in complex systems, 
and it is necessary to take into account the processes
due to the interaction between particles \cite{kabanov2008}.

We write the two-particle interaction as:
\begin{equation}
\label{H_int_pair}
\hat H_{\mathrm{int,pair}} =
 \sum \limits_{\bf k p q } U({\bf q})
{ \hat a^{\dagger}_{\bf k} \hat a^{\dagger}_{\bf p} 
  \hat a_{{\bf p}+{\bf q}} \hat a_{{\bf k}-{\bf q}} 
}
\end{equation}

Next, for convenience, we denote the momenta
of four particles involved in the interaction
${\bf k}$, ${\bf p}$, ${\bf p}+{\bf q}$, ${\bf k}-{\bf q}$ 
as ${\bf 1}$, ${\bf 2}$, ${\bf 3}$, ${\bf 4}$, correspondingly,
and show the conservation law explicitly with Kronecker delta:
\begin{equation}
\label{H_int_pair_1234}
\hat H_{\mathrm{int,pair}} =
 \sum \limits_{\bf 1 2 3 4 } U_{{\bf 3}-{\bf 2}}
{ \hat a^{\dagger}_{\bf 1} \hat a^{\dagger}_{\bf 2}
 \hat a_{\bf 3} \hat a_{\bf 4} 
 \delta_{{\bf 1}+{\bf 2},{\bf 3}+{\bf 4}}
}
\end{equation}

Then, kinetic equations take the form:
\begin{widetext}
\begin{eqnarray}
\label{integer_dn_dt_bose}
\frac{dn_{\bf 1}}{dt}= 
\sum \limits_{\bf 2 3 4 } |U_{{\bf 3}-{\bf 2}}|^2
\left[
(n_{\bf 1}+1) (n_{\bf 2}+1+\delta_{\bf 12}) n_{\bf 3} (n_{\bf 4}-\delta_{\bf 34})
\right.
-
\left.
n_{\bf 1} (n_{\bf 2}-\delta_{\bf 12}) (n_{\bf 3}+1) (n_{\bf 4}+1+\delta_{\bf 34}) 
\right]
 \delta_{\Delta \varepsilon, 0} \delta_{{\bf 1}+{\bf 2},{\bf 3}+{\bf 4}}
\end{eqnarray}
for Bose statistics and
\begin{eqnarray}
\label{integer_dn_dt_fermi}
\frac{dn_{\bf 1}}{dt}= 
\sum \limits_{\bf 2 3 4 } |U_{{\bf 3}-{\bf 2}}|^2
\left[
(1-n_{\bf 1}) (1-n_{\bf 2}-\delta_{\bf 12}) n_{\bf 3} (n_{\bf 4}-\delta_{\bf 34})
\right.
-
\left.
n_{\bf 1} (n_{\bf 2}-\delta_{\bf 12}) (1-n_{\bf 3}) (1-n_{\bf 4}-\delta_{\bf 34}) 
\right]
 \delta_{\Delta \varepsilon, 0} \delta_{{\bf 1}+{\bf 2},{\bf 3}+{\bf 4}}
\end{eqnarray}
for Fermi statistics, respectively.
\end{widetext}

We should note the corrections using $\delta_{\bf 12}$, $\delta_{\bf 34}$
in the terms with coinciding momenta.
While they are negligible in the continual case,
for a finite system the exact form is essential.

As we mentioned earlier,
the practical use of the expressions 
\eqref{integer_dn_dt_bose}, \eqref{integer_dn_dt_fermi}
on the momentum lattice $L \times L \times L$
is hindered by the 
necessity to calculate a large number of terms
which can be estimated as $\sim V^4=L^{12}$,
and as a result, the simulation of system with lattice size 
$L > 8$ is practically impossible \cite{IWOCL2017,ICMSquare_2017}.
However, in the model of the form \eqref{H_int_pair}, 
when the total momentum is conserved,
we can use the analytical transformation 
working in the extended space 
$({\bf k}, \varepsilon)$ of size 
$L^3 \times N_\varepsilon \sim L^5$ 
that dramatically reduces the amount of calculations
to $ \sim$\,$L^5 \ln{L}$.
It allows to greatly increase the affordable system sizes.

\paragraph{Free gas case.}

We demonstrate the principle of this transformation
using the example of a free gas of Fermi or Bose particles
with the energy spectrum $ \varepsilon_{\bf k} \sim {\bf k}^2$
and the contact interaction
$U_{\bf q} = const = U_0$:
\begin{equation}
\nonumber
\hat H_{\mathrm{int,pair}} =
 U_0 \sum \limits_{\bf 1 2 3 4 }
{ \hat a^{\dagger}_{\bf 1} \hat a^{\dagger}_{\bf 2}
 \hat a_{\bf 3} \hat a_{\bf 4} }
 \delta_{{\bf 1}+{\bf 2},{\bf 3}+{\bf 4}}
\end{equation}

First of all, we note that due to Kronecker delta
$\delta_{{\bf 1}+{\bf 2},{\bf 3}+{\bf 4}}$,
the expressions
\eqref{integer_dn_dt_bose}, \eqref{integer_dn_dt_fermi}
have the form of a discrete convolution
\begin{equation}
\label{convolution}
\sum \limits_{k'=1}^{L} f(k') g(k-k') \equiv (f*g)(k),
\end{equation}
which can be efficiently calculated using the Convolution Theorem:
\begin{equation}
\label{convolution_theorem}
(f*g)(k) = \frac{1}{L} \FFT^{-1} \left[ F(r) \cdot G(r) \right] (k),
\end{equation}
where $F(r)$, $G(r)$ are 
the Fourier transforms of functions $f(k)$, $g(k)$.
This allows us to calculate the sum of \eqref{convolution}
with $ \sim L \ln L$ operations
using the Fast Fourier transform (FFT).
In the case of a space of dimension $d$, 
the number of operations is $\sim V \ln V \sim L^d \ln L$.

Second, the same transformation can be performed on the energy axis, since
the discrete particle energies in this model are proportional to integer numbers:
$\varepsilon_n = n \varepsilon_1$, where $n=0 \dots N_\varepsilon-1$.

To correctly use this scheme for non-periodic functions,
the extended range of values $n = 0 \dots N_{\mathrm max}-1$
is employed, with the so-called 'zero-padding' \cite{DSP_Guide}
at $n \ge N_\varepsilon$.

Finally, we introduce functions in the extended space
$({\bf k}, \varepsilon) \equiv {\bf \rho}$ 
and $({\bf r}, \gamma) \equiv {\bf R}$:
\begin{eqnarray}
\label{n_ke}
n_{{\bf k} \varepsilon} \equiv n_{\bf k} \delta_{\varepsilon,\varepsilon_{\bf k}},
\\
\label{s_ke}
s_{{\bf k} \varepsilon} \equiv \delta_{\varepsilon,\varepsilon_{\bf k}},
\\
\label{n_rg}
N_{\bf R} \equiv \FFT ( n_{\bf \rho} ),
\\
\label{s_rg}
S_{\bf R} \equiv \FFT ( s_{\bf \rho} ).
\end{eqnarray}

Substituting them in the expressions 
\eqref{integer_dn_dt_bose}, \eqref{integer_dn_dt_fermi}
and replacing Kronecker delta symbols with sums
$\delta_{\Delta {\bf k}, 0} =
\frac{1}{L^3} \sum \limits_{\bf r} e^{i {\Delta \bf kr}}$ and
$\delta_{\Delta \varepsilon, 0} =
\frac{1}{N_\varepsilon} \sum \limits_\gamma e^{i \Delta \varepsilon \gamma}$,
we obtain the final equation
(the detailed derivation is given in Appendix):
\begin{equation}
\label{pair_finally}
\frac{dn_{\bf k}}{dt}= U_0^2 \left(
[ p_{{\bf k},\varepsilon_{\bf k}}
 + \widetilde p_{2{\bf k},2\varepsilon_{\bf k}} ]
+ n_{\bf k} 
[ q_{{\bf k},\varepsilon_{\bf k}}
 + \widetilde q_{2{\bf k},2\varepsilon_{\bf k}} ]
 \right),
\end{equation}
where
\begin{eqnarray}
\label{finally_2}
p_{\bf \rho} \equiv \FFT^{-1}( P_{\bf R} ),
\\
\nonumber
q_{\bf \rho} \equiv \FFT^{-1}( Q_{\bf R} ),
\\
\nonumber
\widetilde p_{\bf \rho} \equiv \FFT^{-1}( \widetilde P_{\bf R} ),
\\
\nonumber
\widetilde q_{\bf \rho} \equiv \FFT^{-1}( \widetilde Q_{\bf R} ).
\end{eqnarray}

The four-dimensional matrices  
 $P$, $Q$, $\widetilde P$, $\widetilde Q$ in the case of Bose statistics
are given by the expressions:
\begin{widetext}
\begin{eqnarray}
\label{pq_bose}
P_{\bf R} = 
  S_{- \bf R} N^2_{\bf R}
 + 
  N_{- \bf R} N^2_{\bf R}
 - 
  N_{- \bf R} N_{2 \bf R}
 - 
  S_{- \bf R} N_{2 \bf R}
 -
  N_{\bf R} \mathbb{Z}_{\bf R},
\\
\nonumber
\widetilde P_{\bf R} = N^2_{\bf R},
\\
\nonumber
Q_{\bf R} = 
  S_{ - \bf R} n^2_{\bf R}
 - 
  2 N_{ - \bf R} N_{\bf R} S_{\bf R}
 - 
  N_{- \bf R} S^2_{\bf R}
 - 
  N_{- \bf R} S_{2 \bf R}
 - 
  S_{- \bf R} N_{2 \bf R}
 - 
  2 N_{- \bf R} N_{2 \bf R}
 +
  S_{\bf R} \mathbb{Z}_{\bf R},
\\
\nonumber
\widetilde Q_{\bf R} = 
 S^2_{\bf R}
 +
 2 N_{\bf R} S_{\bf R}
 +
 2 N^2_{\bf R}.
\end{eqnarray}
and in the case of Fermi statistics (paying additional attention to spin indices):
\begin{eqnarray}
\label{pq_fermi}
P_{\bf R} = 
  S_{- \bf R} N^2_{\bf R}
 -
  N_{- \bf R} N^2_{\bf R}
 + 
  N_{- \bf R} N_{2 \bf R}
 - 
  S_{- \bf R} N_{2 \bf R}
 +
  N_{\bf R} \mathbb{Z}_{\bf R},
\\
\nonumber
\widetilde P_{\bf R} = - N^2_{\bf R},
\\
\nonumber
Q_{\bf R} = 
 -
  S_{ - \bf R} N^2_{\bf R}
 + 
  2 N_{ - \bf R} N_{\bf R} S_{\bf R}
 - 
  N_{- \bf R} S^2_{\bf R}
 + 
  N_{- \bf R} S_{2 \bf R}
 + 
  S_{- \bf R} N_{2 \bf R}
 - 
  2 N_{- \bf R} N_{2 \bf R}
 -
  S_{\bf R} \mathbb{Z}_{\bf R},
\\
\nonumber
\widetilde Q_{\bf R} = 
 S^2_{\bf R}
 -
 2 N_{\bf R} S_{\bf R}
 +
 2 N^2_{\bf R}.
\end{eqnarray}
The function $\mathbb{Z}_{\bf R}$ is defined in Appendix.
\end{widetext}

As we see, multiple sums 
in the expressions \eqref{integer_dn_dt_bose}, \eqref{integer_dn_dt_fermi}
are converted to several more performance-efficient Fourier transforms.
The achieved reduction in the amount of calculations
makes it possible to increase the available size of the system
to a relatively macroscopic $L \sim 16 \div 48$.
Later in the Chapter \ref{time_results}, we show the
application of this method 
to the detailed study of relaxation times
in weakly interacting Bose and Fermi systems.

Note that more complex models with 
momentum-dependent interaction 
$U_{\bf q} \neq const$ 
can also use this transformation in almost identical way
but the expressions are slightly more complicated.
For example, the second term in the Eq.
\eqref{pq_bose} takes the form:
\begin{equation}
U^2_0 n_{- \bf R} n^2_{\bf R}
\;
\to
\;
 N_{{\bf r} \gamma} \cdot \sum \limits_{{\bf r}'}
  { N_{{\bf r}',-\gamma}   N_{-{\bf r}',\gamma}   
   u_{{\bf r}' + {\bf r}} }
\end{equation}
where $u_{\bf r} \equiv \FFT( U^2_{\bf q} )$.
The sum can again be recognized as a convolution
and calculated with appropriate sequence of Fourier transforms.

\paragraph{The generalization to arbitrary energy levels.}

Now we show how to extend this approach
to the case of arbitrary energy levels $\varepsilon_{\bf k}$,
i.e. for systems with non-parabolic dispersion law
(electrons far from the band edge, 
exciton polaritons, Bogoliubov quasiparticles, etc.).
If the discrete energy levels 
are not proportional to integer numbers,
the precise Fourier transform on the energy axis used in transformation 
\eqref{pq_bose}, 
\eqref{pq_fermi} 
is not possible.

Using the smaller step on the energy axis
we can make the grid values
closer to the actual single-particle energy levels.
The necessary accuracy of the approximation
is determined by the width of the levels
in the physical problem.

In the equations \eqref{integer_dn_dt_bose}, \eqref{integer_dn_dt_fermi} 
we replace 
$\delta_{\Delta \varepsilon, 0}$ to broadening factor
$f(\varepsilon_1 + \varepsilon_2 - \varepsilon_3 - \varepsilon_4)$,
which describes the accuracy of the energy conservation.
Depending on the problem under consideration,
it can be taken in the form of Lorentzian or Gaussian function
\cite{Ando1982}.
The introducing of the broadening factor
alongwith the reasonable choice of smaller energy step
allow us to consider systems 
with an arbitrary not equidistant single-particle spectrum.

We can notice that typical terms
that have received the additional multiplier
(broadening factor)
can still be represented by multiple convolution:
\begin{eqnarray}
\sum \limits_{ \substack{ {\bf k}_2 {\bf k}_3 {\bf k}_4 \\
 \varepsilon_2 \varepsilon_3 \varepsilon_4 }} 
(\dots)
\delta_{\Delta \varepsilon,0}
\to 
\sum \limits_{ \substack{ {\bf k}_2 {\bf k}_3 {\bf k}_4 \\
 \varepsilon_2 \varepsilon_3 \varepsilon_4 }} 
(\dots)
f(\Delta \varepsilon).
\label{bro_dn_dt_bose}
\end{eqnarray}
As a result, the expressions in \eqref{finally_2} become
\begin{eqnarray}
\label{finally_3}
p_{{\bf k}\varepsilon} \equiv 
  \FFT^{-1}( P_{{\bf r}\gamma} F(\gamma) ) ,
\\
\nonumber
q_{{\bf k}\varepsilon} \equiv 
  \FFT^{-1}( Q_{{\bf r}\gamma} F(\gamma) ),
\\
\nonumber
\widetilde p_{{\bf k}\varepsilon} \equiv
  \FFT^{-1}( \widetilde P_{{\bf r}\gamma} F(\gamma) ),
\\
\nonumber
\widetilde q_{{\bf k}\varepsilon} \equiv
  \FFT^{-1}( \widetilde Q_{{\bf r}\gamma} F(\gamma) ),
\end{eqnarray}
where $F(\gamma)$ is the Fourier transform of the factor $f(\varepsilon)$.
   
To verify the correctness of the transformation,
the results of calculation using Eqs.\eqref{pair_finally}--\eqref{finally_3} 
were compared with the direct summation of the original expressions
\eqref{integer_dn_dt_bose}, \eqref{integer_dn_dt_fermi}
for several small systems with dimensions
 4$\times$4$\times$4, 8$\times$8$\times$8
and various particle statistics.
The resulting numbers were equal
with the precision of at least 13 digits 
for all the problem parameters,
proving the validity of the presented method.

As a conclusion, 
in this Chapter we reported the universal approach 
for numerical simulation of the kinetics
in interacting quantum system
with arbitrary energy spectrum
on a finite momentum lattice of relatively large dimensions.

\section{Relaxation time in finite Fermi and Bose systems}
\label{time_results}

Generally, the study of kinetics by numerical solution of equations
\eqref{dn_dt_phonon},\eqref{integer_dn_dt_bose},\eqref{integer_dn_dt_fermi}
is quite laborous.
For problems with occupation numbers near the equilibrium
$n_{\bf k} \simeq n^{(0)}_{\bf k} \equiv f( \varepsilon_{\bf k}, \mu, T)$,
a simple model
based on characteristic relaxation times is often used:
\begin{equation}
\label{tau_model}
\frac{dn_{\bf k}}{dt} = - \frac{1}{\tau_{\bf k}} 
\left( n_{\bf k} - n^{(0)}_{\bf k} \right).
\end{equation}
Values of $\tau_{\bf k}$ are estimated
from 
existing analytical expressions
or experimental data.

There exists a well-known expression 
in the Fermi-liquid theory \cite{kabanov2008}
for the lifetime of quasiparticles near Fermi surface:
\begin{equation}
\label{fermi_liquid_result}
\tau(\varepsilon) \sim \frac{1}{ (\pi k_B T)^2 + (\varepsilon - F)^2 },
\end{equation}
where $F$ is the Fermi level.

For the Bose gas, 
analytical results for lifetime of quasiparticles
in some limit cases are available,
namely, the so-called Beliaev damping \cite{Beliaev1958,Giorgini1998}
 and Landau damping \cite{Giorgini1998,Pitaevskii1997}.

Detailed data on the momentum dependency of characteristic times $\tau_{\bf k}$
would make it possible to improve the quality of model \eqref{tau_model}.
Using numerical calculation
with the method reported in Chapter \ref{lattice_chapter},
we can obtain the values of relaxation times
for the complete range of momenta in the system.
In this Chapter, we present the calculation details
and the results for relaxation times
of weakly interacting Fermi and Bose gases
on a finite momentum lattice.

\begin{figure}
\centering
\includegraphics[width=1.0\linewidth]{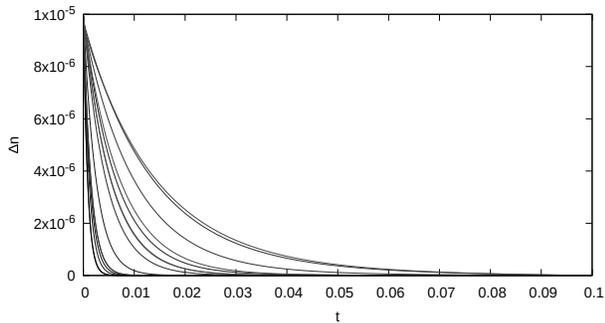}
\caption{Relaxation
of nonequilibrium occupation $ \Delta n_{\bf k}(t)$
for Fermi gas on the lattice $8 \times 8 \times 8$, 
Fermi energy $F=2.0$,
Fermi surface radius $k_F=\pi/2a$.
}
\label{dnk_vs_t}
\end{figure}

\paragraph{Calculation details.}

In order to determine the momentum dependency of relaxation time
$\tau_{\bf k}$, the evolution of the equilibrium system
excited at the corresponding momenta was simulated.
The excitation at the given momentum ${\bf k}$ was done 
 by increasing the occupation number $n_{\bf k}$
by a small value $ \Delta n_{\bf k}(t=0) \sim 10^{-3}$.

The typical time dependency of nonequilibrium occupations
$\Delta n_{\bf k}(t)$ 
demonstrating the exponential-like decay
is shown in Figure \ref{dnk_vs_t}.
The dependency
$\Delta n_{\bf k}(t) = A e^{-t/\tau_{\bf k}}$
allows to use the relation 
for the relaxation time
\begin{equation}
\tau_{\bf k} = -\frac{d n_{\bf k}/dt}{d^2 n_{\bf k}/dt^2},
\end{equation}
where the derivatives are calculated using a finite-difference scheme.

The scale of relaxation time  is determined by energy units. 
In this Chapter, we put $U_0=1$.

\paragraph{Relaxation time in Fermi gas.}

First we calculate the relaxation times $\tau_{\bf k}$
in the weakly interacting Fermi gas
where analytical result \eqref{fermi_liquid_result} exists, 
so that the verification of the simulation method can be made.

The Hamiltonian of the system is:
\begin{equation}
\label{H_results_fermi}
\hat H =
\sum \limits_{{\bf k}\sigma} 
\varepsilon_{\bf k} \hat n_{{\bf k}\sigma}
+ U_0
 \sum \limits_{ \bf k p q }
{ \hat a^{\dagger}_{{\bf k}\uparrow} 
   \hat a^{\dagger}_{{\bf p}\downarrow}
    \hat a_{{\bf p}+{\bf q}\downarrow} \hat a_{{\bf k}-{\bf q}\uparrow} },
\end{equation}
where 
$ \varepsilon_{\bf k} =  \frac{\varepsilon_1}{\Delta k^2} {\bf k}^2$,
$\Delta k=\frac{2 \pi}{La}$ 
is the discreteness of the momentum in the Brillouin zone,
determined by the size of the crystal.

\begin{figure}
\centering
\includegraphics[width=1.0\linewidth]{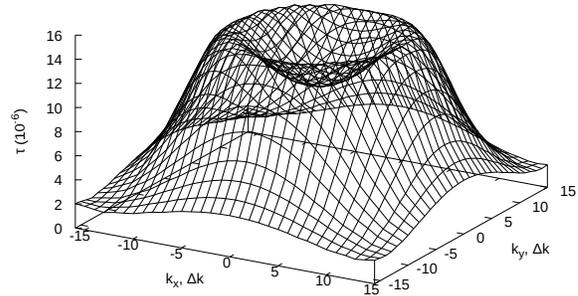}
\caption{Relaxation time
$\tau_{\bf k}$ calculated for
the Fermi gas on the lattice $32 \times 32 \times 32$,
$F=2.0$, 
$k_F=\pi/2a$, $k_B T = 0.7$,
as a function of momentum ${\bf k}=(k_x, k_y, 0)$.
}
\label{fermi_tau_3d}
\end{figure}

\begin{figure}
\centering
\includegraphics[width=1.0\linewidth]{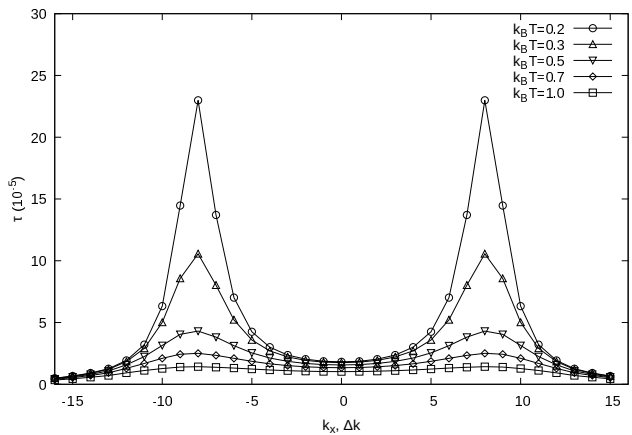}
\caption{Relaxation time 
as a function of momentum ${\bf k}=(k_x, 0, 0)$,
calculated for 
the Fermi gas on the lattice $32 \times 32 \times 32$,
$F=2.0$, 
$k_F=\pi/2a$,
for several temperatures.
Lines are to guide the eye.
}
\label{fermi_many_temp}
\end{figure}

\begin{figure}
\centering
\includegraphics[width=1.0\linewidth]{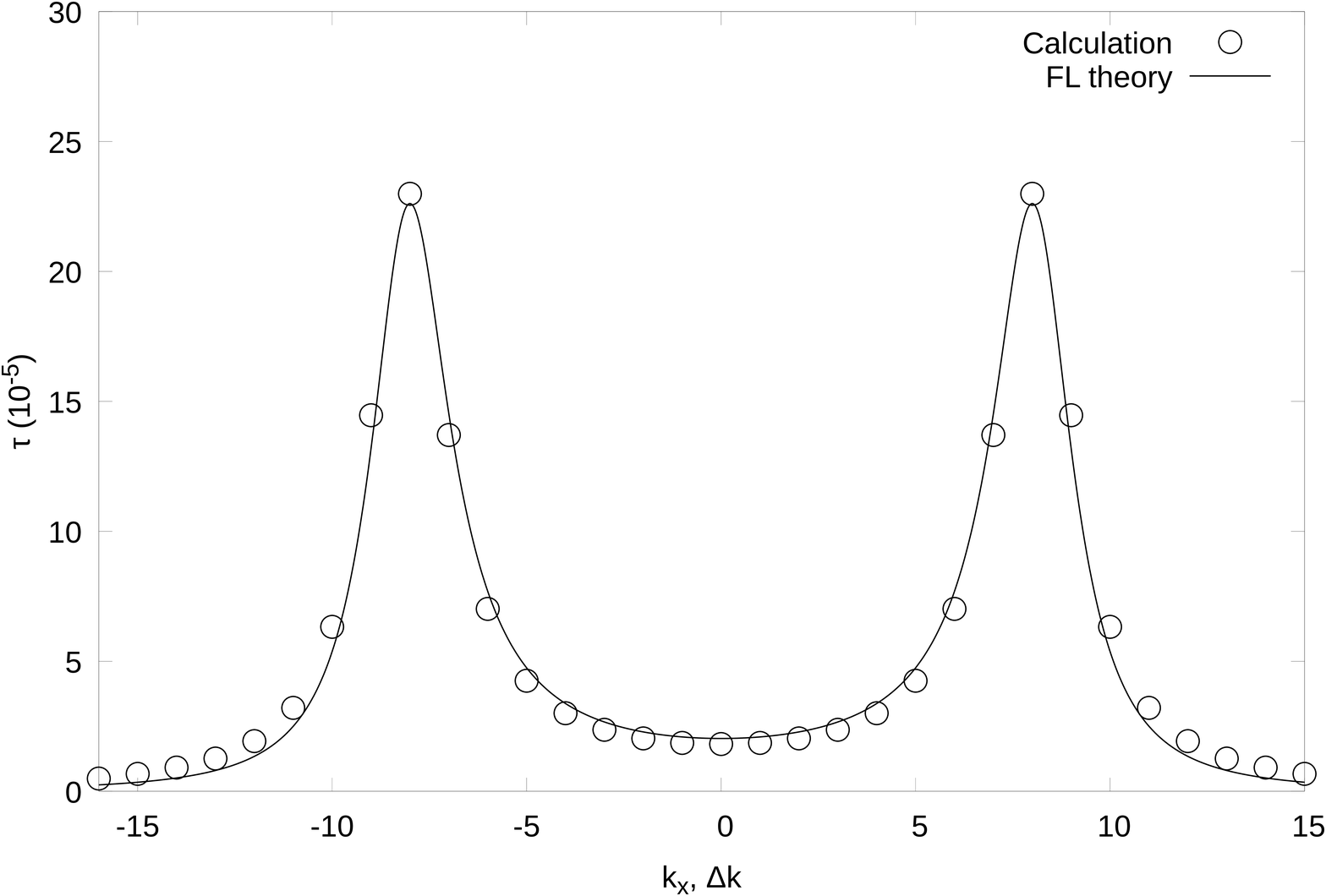}
\caption{The calculated relaxation time
compared with the Fermi-liquid theory, Eq. \eqref{fermi_liquid_result}.
The parameters of the system are the same as for Figure \ref{fermi_many_temp},
temperature $k_B T=0.2$.
}
\label{fermi_vs_theory}
\end{figure}

In the Figures \ref{fermi_tau_3d}, \ref{fermi_many_temp}
the calculated
relaxation time is shown as a function of momentum 
for the Fermi system
at several temperatures
$k_B T=0.2$, $0.3$, $0.5$, $0.7$, $1.0$.
We see the typical Fermi-liquid behaviour
reaching maximum values near the Fermi surface
and proportional to $T^{-2}$.
To demonstrate the overall agreement with the theory,
the values calculated for temperature $k_B T = 0.2$
are shown in the Figure \ref{fermi_vs_theory}
alongwith the analytical dependence \eqref{fermi_liquid_result}.

We can conclude that the application of the reported method
allows to calculate the momentum dependency of the relaxation time
in the weakly interacting quantum gases.

\paragraph{Relaxation time in Bose gas.}

Next we apply this approach to the weakly interacting Bose gas.
The Hamiltonian is
\begin{equation}
\label{H_results_bose}
\hat H =
\sum \limits_{{\bf k}} 
\varepsilon_{\bf k} \hat n_{\bf k}
+ U_0
 \sum \limits_{ \bf k p q }
{ \hat a^{\dagger}_{{\bf k}} 
   \hat a^{\dagger}_{{\bf p}}
    \hat a_{{\bf p}+{\bf q}} \hat a_{{\bf k}-{\bf q}} }.
\end{equation}

\begin{figure}
\begin{center}
\includegraphics[width=1.0\linewidth]{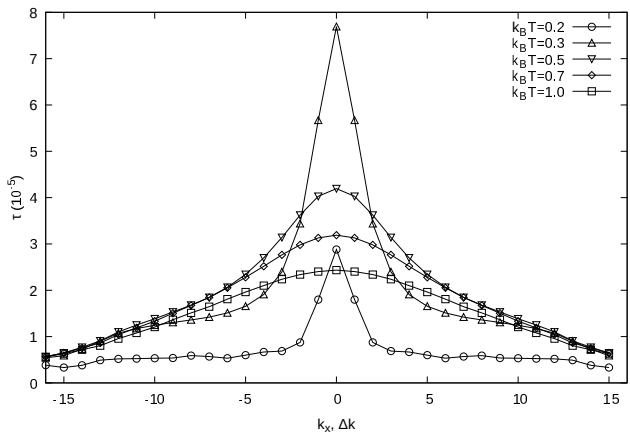}
\end{center}
\caption{Relaxation time 
as a function of momentum  ${\bf k}=(k_x, 0, 0)$,
calculated for 
the Bose gas on the lattice $32 \times 32 \times 32$,
with total particle number $N=10^3$,
for several temperatures
$k_B T=0.2$, $0.3$, $0.5$, $0.7$, $1.0$.
}
\label{bose_many_temp}
\end{figure}

\begin{figure}
\begin{minipage}{1.0\linewidth}\begin{center}
\includegraphics[width=1.0\linewidth]{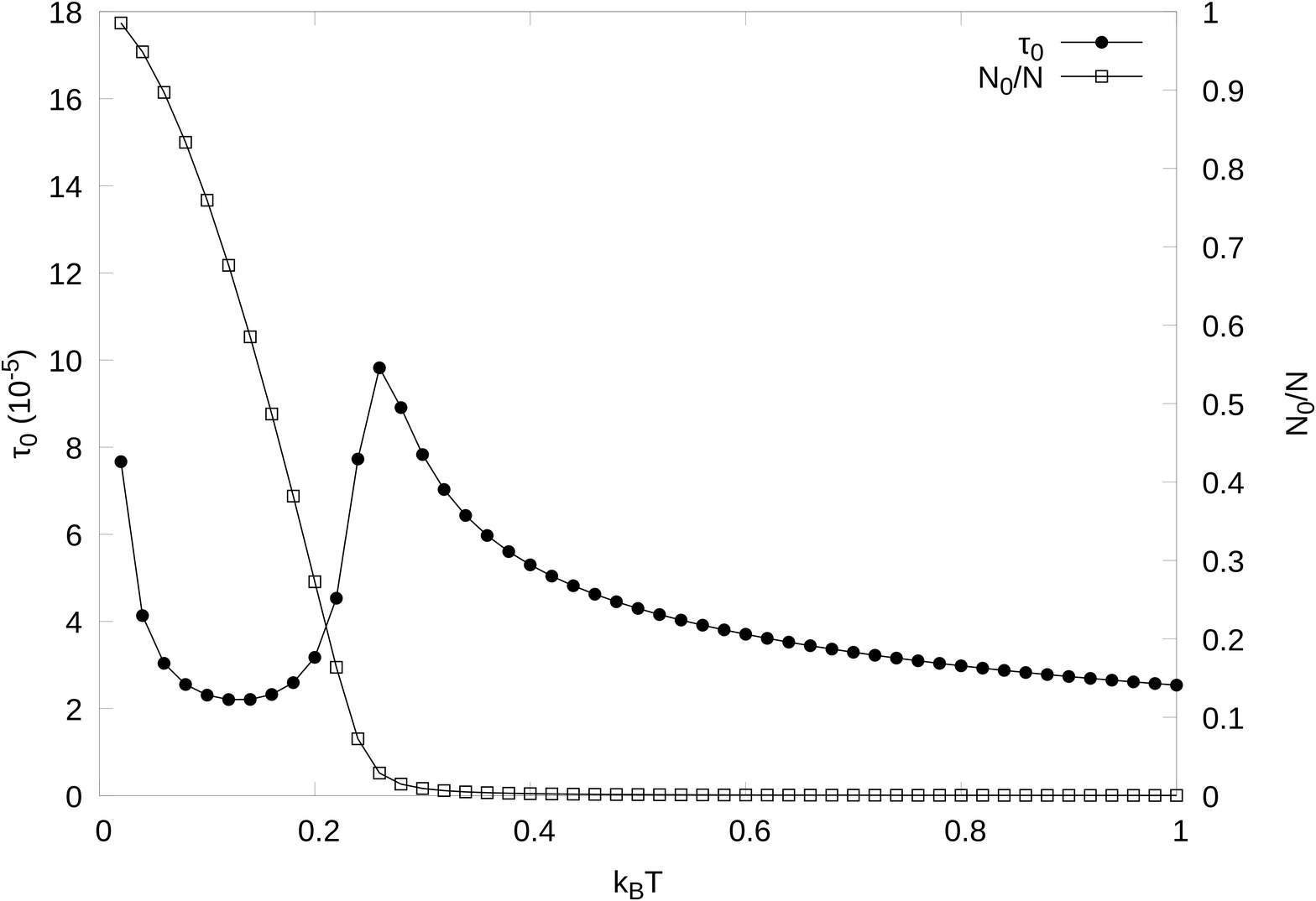}

(a)
\end{center}
\end{minipage}

\begin{minipage}{1.0\linewidth}\begin{center}
\includegraphics[width=1.0\linewidth]{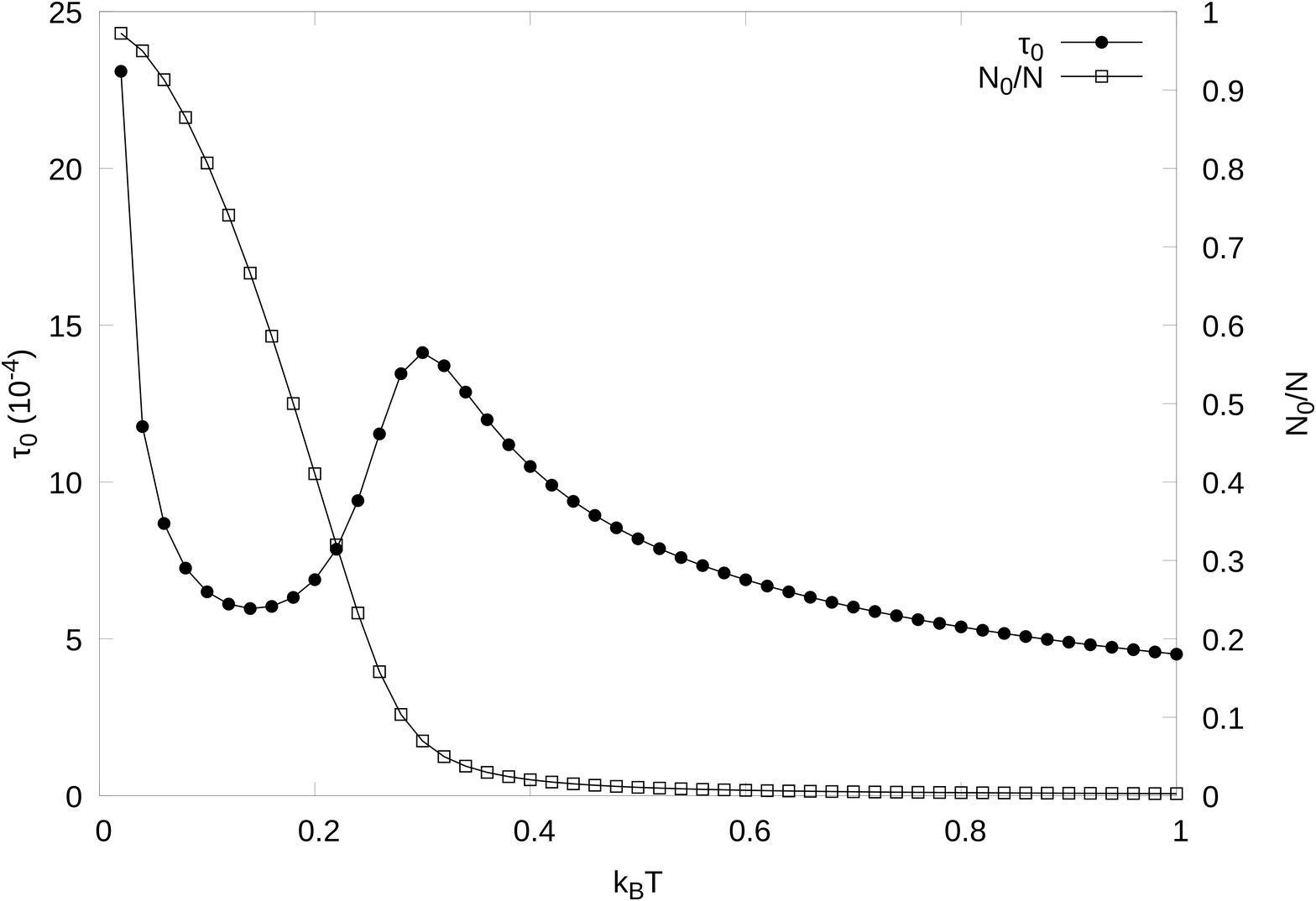}

(b)

\end{center}
\end{minipage}

\begin{minipage}{1.0\linewidth}\begin{center}
\includegraphics[width=1.0\linewidth]{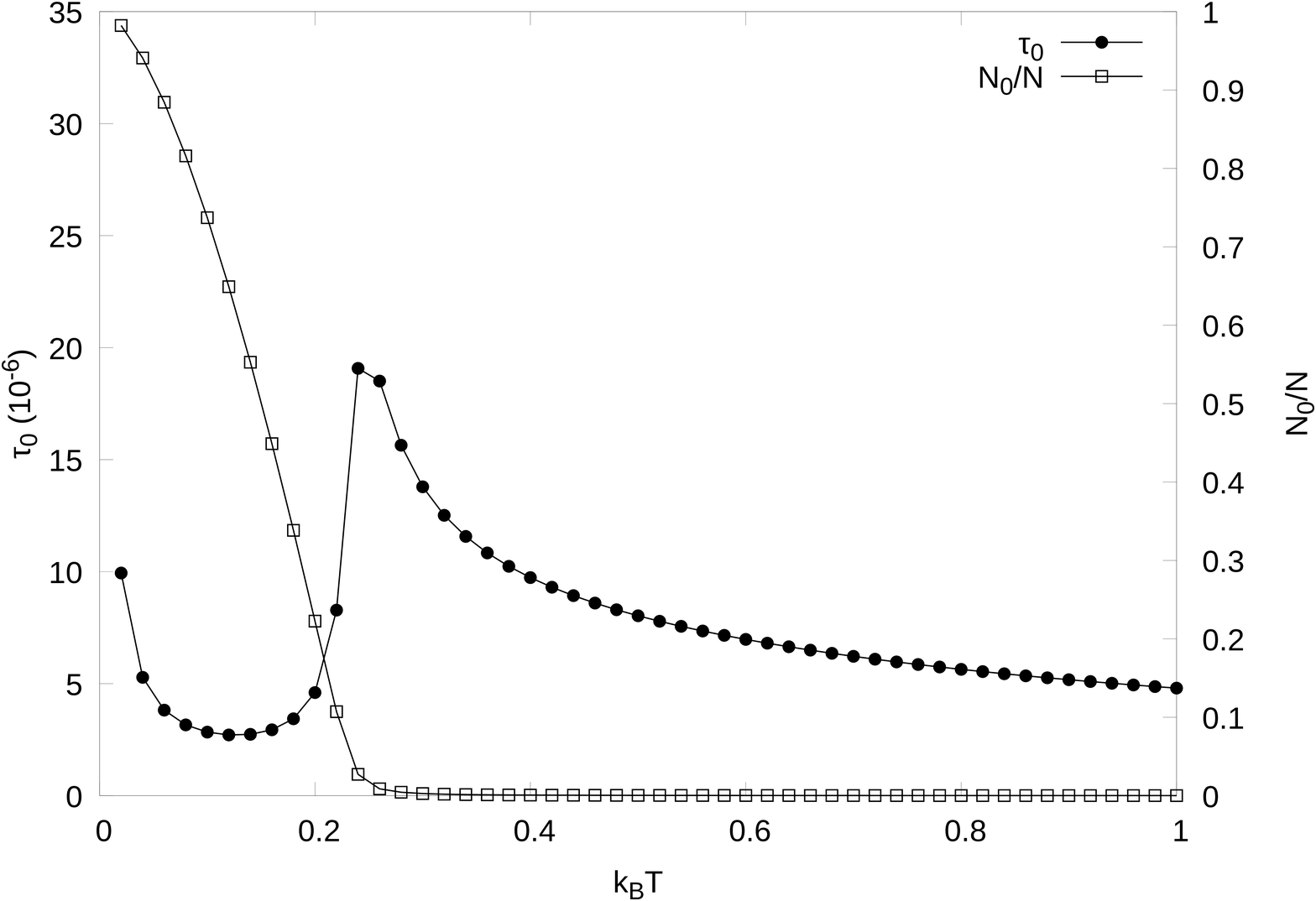}

(c)
\end{center}
\end{minipage}

\caption{Relaxation time 
in Bose gas 
for momentum ${\bf k}=0$
as a function of temperature,
calculated for various lattice sizes $L$=32 (a), 16 (b), 48 (c),
with the particle density large enough to show the BEC transition
(total particle number $N$=1000, 125, 3375, correspondingly).
For reference, the occupation $N_0(T)$ is plotted.
}
\label{size_compare}
\end{figure}


The calculated relaxation times as a function of momentum 
for Bose system with the total particle number N=10$^3$ 
is shown in Figure
\ref{bose_many_temp}.
We see that the relaxation becomes slower 
as the energy $\varepsilon_{\bf k}$ decreases.

The temperature dependency of the relaxation time $\tau_0$ 
at the central momentum ${\bf k}=0$ 
is given in Figure \ref{size_compare} (a).
The parameters of the problem were chosen 
such that the Bose-Einsten condensation
occurs at high enough temperature ($T_c$=0.2236).
The dependence $\tau_0(T)$ shows the pronounced peak
near the transition point.
At low temperatures the second feature 
 is visible.
We associate it to the 
increasing difficulty of the energy transfer
to the discrete spectrum
caused by 
the finite momentum lattice.

To reveal the effect of lattice size, 
we performed the calculations
for different sizes $L$=16, 48 with the same particle density 
($N$=125, 3375, correspondingly).
The functions $\tau_0(T)$ are shown
 in Figures \ref{size_compare} (b) and (c).
As we see, 
the low temperature feature in relaxation time
becomes relatively weaker
with increasing system size,
while the qualitative behaviour 
at large temperatures $T \gtrsim T_c$
remains almost the same
with minor change of the peak sharpness
as the phase transition becomes narrower.

\section{Conclusion}

We presented a universal and efficient method
for numerical simulation of kinetics 
of weakly interacting quantum systems 
on a finite momentum lattice,
using the original transformation to improve the efficiency 
and to increase the affordable system sizes.
This approach can be applied to the wide range of models 
of various statistics with arbitrary single-particle spectrum $\varepsilon_{\bf q}$
and two-particle interaction obeying the momentum conservation.
The system size can be as large as $\sim 50^3$,
which can help to obtain properties 
of continual systems using appropriate extrapolation.

As a demonstration,
the momentum dependence of relaxation time
was calculated for the weakly interacting Fermi gas
on the momentum lattice 32$\times$32$\times$32,
in overall agreement with Fermi-liquid theory predictions
(Figure \ref{fermi_vs_theory}).
Similar calculation for weakly interacting Bose gas 
shows the temperature dependence
with features near to the BEC transition point
and at low temperatures due to discrete energy spectrum
(Figure \ref{size_compare}).

The numerical method reported in this article
can be applied to various time-dependent problems,
such as
relaxation processes in superconductors,
behaviour of nonequilibrium carriers in semiconductors \cite{telenkov2015}
and metals \cite{kabanov2020},
kinetics of atomic gases in magneto-optical traps \cite{nature_physics_2012}, {\it etc}.

\begin{acknowledgments}
The work was supported by the Ministry of Science and Higher Education of 
Russian Federation (state assignment project No. 0723-2020-0036).
\end{acknowledgments}

\appendix
\section{Appendix: Transformation}
\label{appendix_transformation}

In this Appendix, we show how to convert equations 
\eqref{integer_dn_dt_bose}, \eqref{integer_dn_dt_fermi}
to
\eqref{pair_finally}, \eqref{pq_bose}, \eqref{pq_fermi}
in the case of Bose statistics.

First, we expand the 
brackets in the equation \eqref{integer_dn_dt_bose}:
\begin{widetext}
\begin{eqnarray}
\label{dn_dt_bose_detailed}
\frac{dn_{{\bf k}_1}}{dt}=
U_0^2
\sum \limits_{ {\bf k}_2 {\bf k}_3 {\bf k}_4 } 
\left\{
[ n_{\bf 2} n_{\bf 3} n_{\bf 4} 
 + n_{\bf 3} n_{\bf 4} 
 + n_{\bf 1} (
	n_{\bf 3} n_{\bf 4} 
	- n_{\bf 2} n_{\bf 3} 
	- n_{\bf 2} n_{\bf 4} 
	- n_{\bf 2} 
]
\right.
 \\
\nonumber
- \delta_{\bf 34} [
  (n_{\bf 2} n_{\bf 3} +  n_{\bf 3}) + n_{\bf 1} (
	 2 n_{\bf 2} n_{\bf 3} + n_{\bf 2} + n_{\bf 3}
  )
]
\\
\nonumber
+ \delta_{\bf 12} [
  n_{\bf 3} n_{\bf 4} + n_{\bf 1} (
	 2 n_{\bf 3} n_{\bf 4} + n_{\bf 3} + n_{\bf 4} + 1
  )
]
 \\
\nonumber
\left.
+ \delta_{\bf 12} \delta_{\bf 34} (n_{\bf 1} - n_{\bf 3})
\right\}
 \delta_{\varepsilon_1+\varepsilon_2, \varepsilon_3+\varepsilon_4} 
  \delta_{{\bf 1}+{\bf 2},{\bf 3}+{\bf 4}}
\end{eqnarray}
\end{widetext}

Here we grouped the terms where the particle momenta coincide 
in the initial or final states
($\delta_{\bf 12}$ and $\delta_{\bf 34}$, correspondingly).


\paragraph{The first line} 
in the expression \eqref{dn_dt_bose_detailed}
can be conveniently written as:
\begin{equation}
\label{bose_detailed_first_line}
\left(\frac{dn_{{\bf k}_1}}{dt}\right)^{(1)}=
U_0^2
\left[
(
A_{{\bf k}_1} + B_{{\bf k}_1}
)
+
 n_{{\bf k}_1}
(
B_{{\bf k}_1} - 2C_{{\bf k}_1} - D_{{\bf k}_1}
)
\right],
\end{equation}
where
\begin{eqnarray}
\label{sum_n234}
A_{{\bf k}_1} \equiv \sum \limits_{ {\bf k}_2 {\bf k}_3 {\bf k}_4 } 
 n_{{\bf k}_2} n_{{\bf k}_3} n_{{\bf k}_4}
  \delta_{\varepsilon_1+\varepsilon_2, \varepsilon_3+\varepsilon_4} 
   \delta_{{{\bf k}_1}+{{\bf k}_2},{{\bf k}_3}+{{\bf k}_4}}, \\
\label{sum_n34}
B_{{\bf k}_1} \equiv \sum \limits_{ {\bf k}_2 {\bf k}_3 {\bf k}_4 } 
 n_{{\bf k}_3} n_{{\bf k}_4}
  \delta_{\varepsilon_1+\varepsilon_2, \varepsilon_3+\varepsilon_4} 
   \delta_{{{\bf k}_1}+{{\bf k}_2},{{\bf k}_3}+{{\bf k}_4}},\\
\label{sum_n23}
C_{{\bf k}_1} \equiv \sum \limits_{ {\bf k}_2 {\bf k}_3 {\bf k}_4 }
 n_{{\bf k}_2} n_{{\bf k}_3}
  \delta_{\varepsilon_1+\varepsilon_2, \varepsilon_3+\varepsilon_4} 
   \delta_{{{\bf k}_1}+{{\bf k}_2},{{\bf k}_3}+{{\bf k}_4}},\\
\label{sum_n2}
D_{{\bf k}_1} \equiv \sum \limits_{ {\bf k}_2 {\bf k}_3 {\bf k}_4 }
 n_{{\bf k}_2}
  \delta_{\varepsilon_1+\varepsilon_2, \varepsilon_3+\varepsilon_4} 
   \delta_{{{\bf k}_1}+{{\bf k}_2},{{\bf k}_3}+{{\bf k}_4}}
.
\end{eqnarray}

We rewrite the expression \eqref{sum_n234}
in the expanded space
$({\bf k}, \varepsilon) \equiv {\bf \rho}$,
employ the notation \eqref{n_ke}-\eqref{s_rg},
and replace Kronecker delta symbols with sums:
\begin{eqnarray}
\label{A_ke_1}
A_{{\bf k}_1 \varepsilon_1} = 
 \frac{1}{L^3 N_\varepsilon} \sum \limits_{ 
 \substack{  {\bf k}_2 {\bf k}_3 {\bf k}_4  \\
 \varepsilon_2 \varepsilon_3 \varepsilon_4} 
} 
 n_{{\bf k}_2 \varepsilon_2} n_{{\bf k}_3 \varepsilon_3} n_{{\bf k}_4 \varepsilon_4}
\\ \nonumber
\times
 \sum \limits_{{\bf r}\gamma}
    e^{i ({{\bf k}_1}+{{\bf k}_2}-{{\bf k}_3}-{{\bf k}_4}) {\bf r} } \,
    e^{i (\varepsilon_1 + \varepsilon_2 - \varepsilon_3 - \varepsilon_4) \gamma }
\end{eqnarray}

Using the relation
$n_{{\bf k} \varepsilon} = \sum\limits_{{\bf r}\gamma} N_{{\bf r} \gamma} 
e^{i  ({\bf k}{\bf r} + \varepsilon \gamma) }$,
we can write:
\begin{eqnarray}
\label{A_ke_2}
A_{{\bf k}_1 \varepsilon_1} = 
 \frac{1}{L^3 N_\varepsilon}
 \sum \limits_{  \substack{  {\bf k}_2 {\bf k}_3 {\bf k}_4  \\
 \varepsilon_2 \varepsilon_3 \varepsilon_4} 
}
 \sum \limits_{  \substack{  {\bf r}_2 {\bf r}_3 {\bf r}_4  \\
 \gamma_2 \gamma_3 \gamma_4} 
} 
 N_{{\bf r}_2 \gamma_2} N_{{\bf r}_3 \gamma_3} N_{{\bf r}_4 \gamma_4}
\\ \nonumber
\times
 \sum \limits_{{\bf r}\gamma}
    e^{i ({\bf k}_2 {\bf r}_2 +{\bf k}_3 {\bf r}_3+{\bf k}_4 {\bf r}_4)}
    e^{i (\varepsilon_2 \gamma_2 + \varepsilon_3 \gamma_3 + \varepsilon_4 \gamma_4)  }
\\ \nonumber
\times
    e^{i ({{\bf k}_1}+{{\bf k}_2}-{{\bf k}_3}-{{\bf k}_4}) {\bf r} } 
    e^{i (\varepsilon_1 + \varepsilon_2 - \varepsilon_3 - \varepsilon_4) \gamma }
\\ \nonumber
= \frac{1}{L^3 N_\varepsilon} 
\sum \limits_{{\bf r}\gamma}
N_{-{\bf r}, -\gamma} N_{{\bf r}\gamma} N_{{\bf r}\gamma} 
e^{i ({\bf k}_1 {\bf r} + \varepsilon_1 \gamma) }, 
\end{eqnarray}
which takes in the form of the Fourier transform.

The functions $B$, $C$, $D$ can be converted in the same way.
As a result, we obtain:
\begin{eqnarray}
\label{A_rg}
A_{{\bf r} \gamma} = N_{-{\bf r}, -\gamma} (N_{{\bf r} \gamma})^2,\\
\label{B_rg}
B_{{\bf r} \gamma} = S_{{-{\bf r}, -\gamma}} (N_{{\bf r} \gamma})^2,\\
\label{C_rg}
C_{{\bf r} \gamma} = N_{{-{\bf r}, -\gamma}} N_{{\bf r} \gamma} S_{{\bf r} \gamma}, \\
\label{D_rg}
D_{{\bf r} \gamma} = N_{{-{\bf r}, -\gamma}} (S_{{\bf r} \gamma})^2,
\end{eqnarray}
where $S_{{\bf r} \gamma}$ is the Fourier transform of the previously introduced function
$s_{{\bf k} \varepsilon} \equiv \delta_{\varepsilon,\varepsilon_{\bf k}}$.

The function $A_{{\bf r} \gamma}$, $\dots$, $D_{{\bf r} \gamma}$ are turned into
$A_{{\bf k} \varepsilon}$, $\dots$, $D_{{\bf k} \varepsilon}$
using the inverse Fourier transform,
after that the desired values $A_{\bf k}$, $\dots$, $D_{\bf k}$ 
for the expression \eqref{bose_detailed_first_line}
can be obtained with $\varepsilon=\varepsilon_{\bf k}$. 


\paragraph{The second line} in the expression \eqref{dn_dt_bose_detailed}
corresponds to the case when ${\bf k}_3={\bf k}_4$.
It can be written as:
\begin{widetext}
\begin{equation}
\label{bose_detailed_line_12}
\left(\frac{dn_{{\bf k}_1}}{dt}\right)^{(2)}
= - U_0^2
\left[
( E_{{\bf k}_1} + F_{{\bf k}_1} )
+
 n_{{\bf k}_1}
(
2 E_{{\bf k}_1} + F_{{\bf k}_1} + G_{{\bf k}_1}
)
\right],
\end{equation}
where
\begin{eqnarray}
\label{sum_34_n23}
E_{{\bf k}_1} \equiv \sum \limits_{ {\bf k}_2 {\bf k}_3 {\bf k}_4 } 
 n_{{\bf k}_2} n_{{\bf k}_3} 
 \delta_{{{\bf k}_3},{{\bf k}_4}}
  \delta_{\varepsilon_1+\varepsilon_2, \varepsilon_3+\varepsilon_4} 
   \delta_{{{\bf k}_1}+{{\bf k}_2},{{\bf k}_3}+{{\bf k}_4}}, \\
\label{sum_n3}
F_{{\bf k}_1} \equiv \sum \limits_{ {\bf k}_2 {\bf k}_3 {\bf k}_4 } 
 n_{{\bf k}_3}
 \delta_{{{\bf k}_3},{{\bf k}_4}}
  \delta_{\varepsilon_1+\varepsilon_2, \varepsilon_3+\varepsilon_4} 
   \delta_{{{\bf k}_1}+{{\bf k}_2},{{\bf k}_3}+{{\bf k}_4}},\\
\label{sum_n2}
G_{{\bf k}_1} \equiv \sum \limits_{ {\bf k}_2 {\bf k}_3 {\bf k}_4 } 
 n_{{\bf k}_2}
 \delta_{{{\bf k}_3},{{\bf k}_4}}
  \delta_{\varepsilon_1+\varepsilon_2, \varepsilon_3+\varepsilon_4} 
   \delta_{{{\bf k}_1}+{{\bf k}_2},{{\bf k}_3}+{{\bf k}_4}}.
\end{eqnarray}
\end{widetext}

A reasoning similar to the previous paragraph gives the expressions:
\begin{eqnarray}
\label{E_rg}
E_{{\bf r} \gamma} = N_{-{\bf r}, -\gamma} N_{2{\bf r} 2\gamma},\\
\label{F_rg}
F_{{\bf r} \gamma} = S_{{-{\bf r}, -\gamma}} N_{2{\bf r} 2\gamma},\\
\label{G_rg}
G_{{\bf r} \gamma} = N_{{-{\bf r}, -\gamma}} S_{2 {\bf r} 2\gamma}.
\end{eqnarray}


\paragraph{The third line} in the expression \eqref{dn_dt_bose_detailed}
corresponds to the case when ${\bf k}_1={\bf k}_2$.
Taking into consideration the same 
role of ${\bf k}_3$ and ${\bf k}_4$ in the sums,
this line can be written as:
\begin{equation}
\label{bose_detailed_line_12}
\left(\frac{dn_{{\bf k}_1}}{dt}\right)^{(3)}
= U_0^2
\left[
H_{{\bf k}_1} 
+
 n_{{\bf k}_1}
(
2 H_{{\bf k}_1} + 2 I_{{\bf k}_1} + J_{{\bf k}_1}
)
\right],
\end{equation}
where

\begin{widetext}
\begin{eqnarray}
\label{sum_12_n34}
H_{{\bf k}_1} \equiv \sum \limits_{ {\bf k}_2 {\bf k}_3 {\bf k}_4 } 
 n_{{\bf k}_3} n_{{\bf k}_4}
 \delta_{{{\bf k}_1},{{\bf k}_2}}
  \delta_{\varepsilon_1+\varepsilon_2, \varepsilon_3+\varepsilon_4} 
   \delta_{{{\bf k}_1}+{{\bf k}_2},{{\bf k}_3}+{{\bf k}_4}}, \\
\label{sum_12_n3}
I_{{\bf k}_1} \equiv \sum \limits_{ {\bf k}_2 {\bf k}_3 {\bf k}_4 } 
 n_{{\bf k}_3}
 \delta_{{{\bf k}_1},{{\bf k}_2}}
  \delta_{\varepsilon_1+\varepsilon_2, \varepsilon_3+\varepsilon_4} 
   \delta_{{{\bf k}_1}+{{\bf k}_2},{{\bf k}_3}+{{\bf k}_4}}, \\
\label{sum_12_1}
J_{{\bf k}_1} \equiv \sum \limits_{ {\bf k}_2 {\bf k}_3 {\bf k}_4 } 
 \delta_{{{\bf k}_1},{{\bf k}_2}}
  \delta_{\varepsilon_1+\varepsilon_2, \varepsilon_3+\varepsilon_4} 
   \delta_{{{\bf k}_1}+{{\bf k}_2},{{\bf k}_3}+{{\bf k}_4}}.
\end{eqnarray}

Repeating the transformation made in \eqref{A_ke_1},
we convert the equation \eqref{sum_12_n34} to the form:
\begin{equation}
\label{H_ke}
H_{{\bf k}_1 \varepsilon_1} = 
 \frac{1}{L^3 N_\varepsilon} 
  \sum \limits_{{\bf r}\gamma}
   N_{{\bf r} \gamma} N_{{\bf r}\gamma} e^{i (2 {\bf k}_1 {\bf r} + 2 \varepsilon_1 \gamma) }
= \tilde H_{2{\bf k}_1, 2\varepsilon_1}
\end{equation}
\end{widetext}
i.e. the required values of 
$H_{{\bf k}_1 \varepsilon_1}$
are calculated 
using the Fourier transform of the auxiliary function
\begin{equation}
\label{Htilde_rg}
\tilde H_{{\bf r} \gamma} = ( N_{{\bf r} \gamma} )^2.
\end{equation}

The auxiliary functions for variables 
$I_{{\bf k}_1 \varepsilon_1}$ and $J_{{\bf k}_1 \varepsilon_1}$
are obtained similarly:
\begin{eqnarray}
\label{Itilde_rg}
\tilde I_{{\bf r} \gamma} = N_{{\bf r} \gamma} S_{{\bf r} \gamma},\\
\label{Jtilde_rg}
\tilde J_{{\bf r} \gamma} = ( S_{{\bf r} \gamma} )^2.
\end{eqnarray}


\paragraph{The fourth line} in the expression \eqref{dn_dt_bose_detailed}
corresponds to the case when ${\bf k}_1={\bf k}_2$ and ${\bf k}_3={\bf k}_4$.
It can be written as:
\begin{equation}
\label{bose_detailed_line_1234}
\left(\frac{dn_{{\bf k}_1}}{dt}\right)^{(4)}
= U_0^2
\left[
- X_{{\bf k}_1} 
+
 n_{{\bf k}_1} Y_{{\bf k}_1}
\right],
\end{equation}
where
\begin{widetext}
\begin{eqnarray}
\label{sum_1234_n3}
X_{{\bf k}_1} \equiv \sum \limits_{ {\bf k}_2 {\bf k}_3 {\bf k}_4 } 
 n_{{\bf k}_3}
 \delta_{{{\bf k}_1},{{\bf k}_2}}
 \delta_{{{\bf k}_3},{{\bf k}_4}}
  \delta_{\varepsilon_1+\varepsilon_2, \varepsilon_3+\varepsilon_4} 
   \delta_{{{\bf k}_1}+{{\bf k}_2},{{\bf k}_3}+{{\bf k}_4}}, \\
\label{sum_1234_1}
Y_{{\bf k}_1} \equiv \sum \limits_{ {\bf k}_2 {\bf k}_3 {\bf k}_4 } 
 \delta_{{{\bf k}_1},{{\bf k}_2}}
 \delta_{{{\bf k}_3},{{\bf k}_4}}
  \delta_{\varepsilon_1+\varepsilon_2, \varepsilon_3+\varepsilon_4} 
   \delta_{{{\bf k}_1}+{{\bf k}_2},{{\bf k}_3}+{{\bf k}_4}}.
\end{eqnarray}
\end{widetext}

Taking into consideration the equalities
${\bf k}_1={\bf k}_2$, ${\bf k}_3={\bf k}_4$,
the relation  ${\bf k}_1+{\bf k}_2={\bf k}_3+{\bf k}_4$
can be rewritten as $2{\bf k}_1 = 2{\bf k}_3 + {\mathbb G}$.
It means that the momenta ${\bf k}_1$ and ${\bf k}_3$ 
are either equal or differ by a half 
of the unit vector of the reciprocal lattice.
To obtain all different vectors ${\bf k}_3$
for a given ${\bf k}_1$,
we should use all half-vectors
of reciprocal lattice (including zero vector)
 contained in the first Brillouin zone. 
We denote them as ${\mathbb G}/2$.

In this case, 
we get the expressions similar to \eqref{A_ke_2}
with extra factor
 $e^{i ({\bf k}_1 - {\bf k}_3) {\bf r}} = e^{i {\bf r \mathbb{G}}/2}$:
\begin{eqnarray}
\label{X_ke}
X_{{\bf k}_1 \varepsilon_1} = 
 \frac{1}{L^3 N_\varepsilon} 
  \sum \limits_{{\bf r}\gamma}
   e^{i ( {{\bf k}_1} {\bf r} + \varepsilon_1 \gamma) }
    N_{{\bf r} \gamma} 
     \sum \limits_{{\bf \mathbb{G}}/2}
       e^{i {\bf r} {\bf \mathbb{G}}/2 }, \\
\label{Y_ke}
Y_{{\bf k}_1 \varepsilon_1} = 
 \frac{1}{L^3 N_\varepsilon} 
  \sum \limits_{{\bf r}\gamma}
   e^{i ( {{\bf k}_1} {\bf r} + \varepsilon_1 \gamma) }
    S_{{\bf r} \gamma} 
     \sum \limits_{{\bf \mathbb{G}}/2}
       e^{i {\bf r} {\bf \mathbb{G}}/2 }.
\end{eqnarray}

The factor 
$
\sum \limits_{{\bf \mathbb{G}/2}} e^{i {\bf r} {\bf \mathbb{G}}/2 }
 \equiv 
\mathbb{Z}_{\bf r}
$ 
contains $2^d$ terms (where $d$ is the dimension of space) 
and is equal to either 0 or $2^d$,
depending on the components of the vector ${\bf r}$.

As a result, we obtain:
\begin{eqnarray}
\label{X_rg}
\tilde X_{{\bf r} \gamma} = N_{{\bf r} \gamma} {\mathbb Z}_{\bf r},\\
\label{Y_rg}
\tilde Y_{{\bf r} \gamma} = S_{{\bf r} \gamma} {\mathbb Z}_{\bf r}.
\end{eqnarray}

The relations for Fermi statistics 
are derived using the same reasoning.
Combining all the terms $A_{{\bf r} \gamma}$, $\dots$, $Y_{{\bf r} \gamma}$,
we get the expressions \eqref{pq_bose}, \eqref{pq_fermi}.

\end{document}